\documentclass[12pt]{iopart}

\usepackage{graphicx}
\begin{document}

\title[Properties of particle production at large transverse momentum...]{Properties of particle production at large transverse momentum in Au+Au and Cu+Cu collisions at RHIC}

\author{Bedangadas Mohanty \footnote{On leave from:
Variable Energy Cyclotron Centre, 1/AF, Bidhan Nagar, Kolkata 700064, India}(for the STAR Collaboration)}

\address{Lawrence Berkeley National Laboratory,
One Cyclotron Road, Berkeley, California 94720}
\ead{bmohanty@lbl.gov}
\begin{abstract}

We present the incident energy and system size dependence 
of the $p_{\mathrm T}$ spectra for $\pi^{\pm}$, $p$, and $\bar{p}$ 
using Au+Au and Cu+Cu collisions at $\sqrt{s_{\mathrm {NN}}}$
= 62.4 and 200 GeV in STAR experiment at RHIC. 
Through these measurements in the $p_{\mathrm T}$ range of 
0.2~$<$~$p_{\mathrm T}$~$<$~10~GeV/$c$ we conduct a systematic 
study of the beam energy, system size and particle species 
dependence of nuclear modification factors and address
specific predictions from the quark coalescence
models regarding the beam energy dependence of baryon 
enhancement in the intermediate $p_{\mathrm T}$ (2 $<$ $p_{\mathrm T}$ $<$ 6 GeV/$c$) region.
\end{abstract}


\section{Introduction}

    A detailed study of the beam energy ($\sqrt{s_{\mathrm {NN}}}$), 
    transverse momentum ($p_{\mathrm T}$) 
    and system size dependence of identified hadron 
    production will provide the necessary data to understand the 
    mechanism of energy loss, and put constraints on parameters in 
    energy loss model calculations
    like initial gluon density~\cite{vitev_density} and life time of dense matter~\cite{xnwang_lifetime}. 
    They will also help in probing the difference in the energy loss of 
    quarks and gluons in the medium~\cite{xnwang_nonabelian,star_pid200}.
    The energy dependence of baryon to meson ratio at the intermediate 
    $p_{\mathrm T}$ region from 2 - 6 GeV/$c$ will address 
    the specific prediction from the quark coalescence models of
    a higher baryon to meson ratio at $\sqrt{s_{\mathrm {NN}}}$~=~62.4~GeV
    compared to $\sqrt{s_{\mathrm {NN}}}$~=~200~GeV~\cite{vitev_62}. 
    At high $p_{\mathrm T}$ ($>$~6~GeV/$c$) the $p$/$\pi$ ratio can provide
    information on quark and gluon jet conversions in the medium formed
    in heavy ion collisions~\cite{ko}.  

\section{Experiment and Analysis}

The data presented here are obtained from the 
Time Projection Chamber (TPC) and the 
Time-Of-Flight (TOF) detector in the STAR 
experiment~\cite{starnim} at RHIC in the year 2004. Measurements 
of the ionization energy loss of charged tracks in 
the TPC gas under a magnetic field of 0.5 Tesla 
are used to identify $\pi^{\pm}$, $p$ and $\bar{p}$ 
within $\mid\eta\mid$~$<$~0.5 and full azimuth, 
for $p_{\mathrm T}~\le~1.1$~GeV/$c$ 
and $2.5~\le~p_{\mathrm T}~\le 10$~GeV/$c$. 
The data from TOF provides particle 
identification up to $p_{T}\sim3$~GeV/$c$ for pions 
and protons in $-1\!<\!\eta\!<\!0$ in pseudorapidity 
and  $\pi/30$ rad in azimuth.
Weak-decay feed-down (e.g. $K_{S}^{0}\rightarrow\pi^{+}\pi^{-}$) to the pion
spectra was subtracted from the results presented.
Inclusive $p$ and $\bar{p}$ production include all of the hyperon
feed-downs to reflect the total baryon production. 
Systematic errors for the TOF measurements are 
similar at both energies and are around 8\%~\cite{ppdau}. 
The total systematic errors on charged pion yields at both energies
and collision systems are estimated to be ${}^<_\sim$ 15\% and those for proton and
anti-protons to be ${}^<_\sim$ 26\% over the entire $p_{\mathrm T}$ 
range studied~\cite{ppdau}. 

\section{Results}

\subsection{Nuclear Modification Factor}

\begin{figure}
\begin{center}
\includegraphics[scale=0.3]{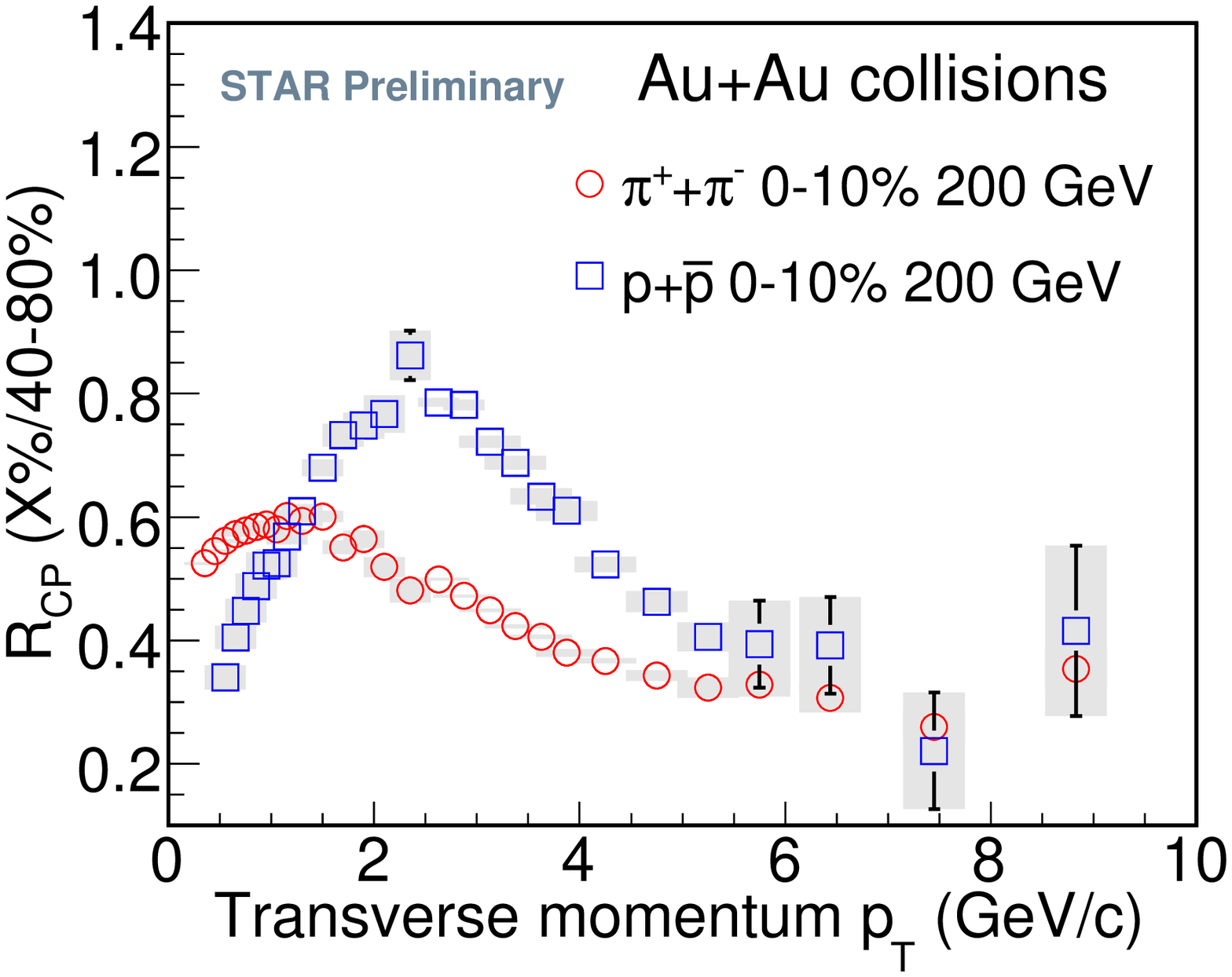}
\includegraphics[scale=0.3]{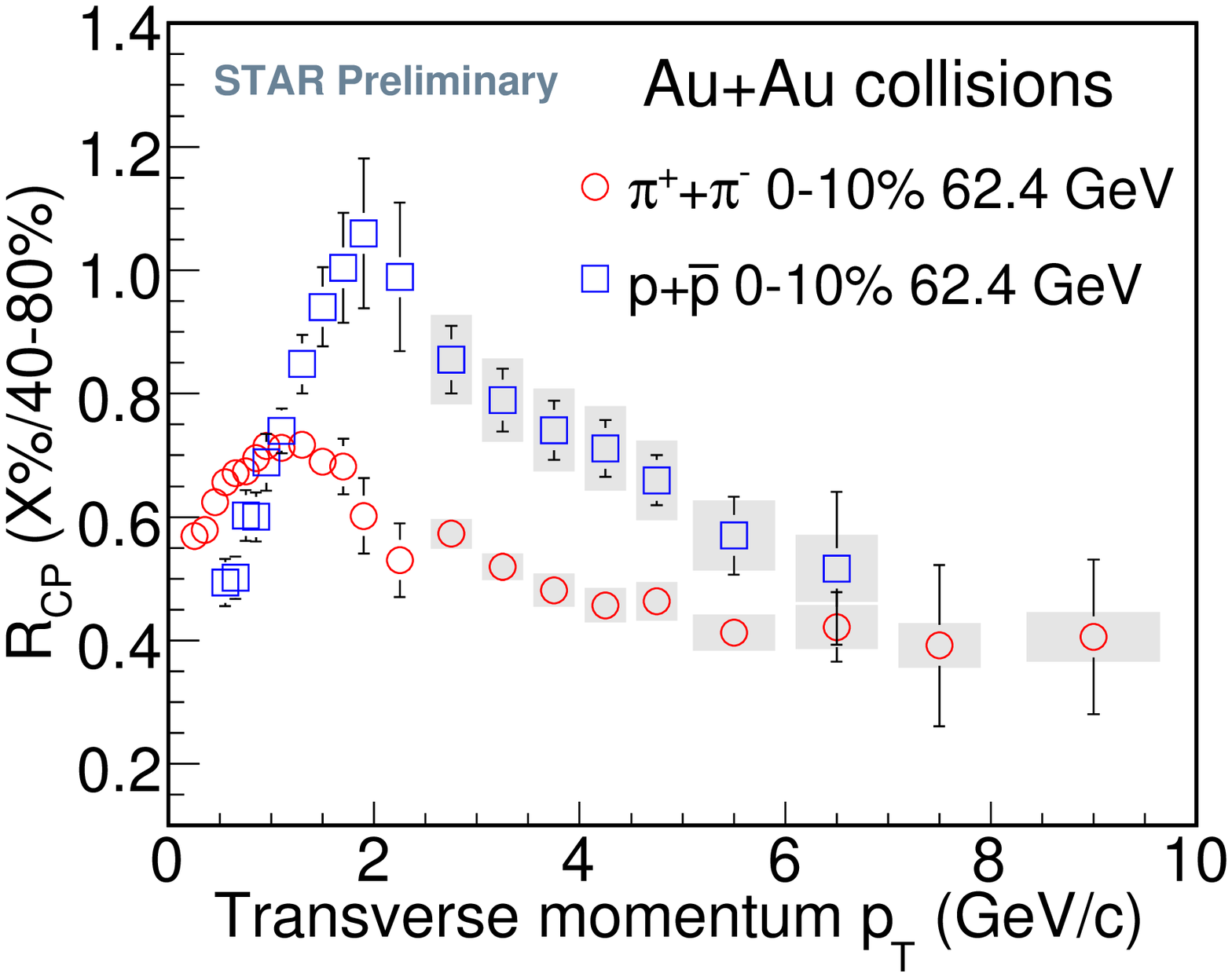}
\caption{$R_{\mathrm {CP}}$ for 
$\pi^{+}$+$\pi^{-}$ and
$p$+$\bar{p}$ in Au+Au collisions at 
$\sqrt{s_{\mathrm {NN}}}$ = 200 (left panel) and 62.4 GeV (right panel).
The error bars are statistical, the shaded bands are systematic errors.
There is an additional error of about 10\% not shown in the figure due 
to uncertainty in $N_{\rm {bin}}$ calculations.}
\label{fig1}
\end{center}
\end{figure}

One nuclear modification factor ($R_{\mathrm {CP}}$) is defined as
$R_{\rm{CP}}(p_{\rm T})\,=\,\frac{\langle N_{\rm {bin}}^{\rm peri}\rangle d^2N_{\rm{cent}}/dy dp_{\rm T}}{\langle N_{\rm {bin}}^{\rm cent}\rangle \,d^2N_{\rm
{peri}}/dy dp_{\rm T}}$,
where $\langle N_{\mathrm {bin}}\rangle$ is the average number
of binary nucleon-nucleon collisions per event.
Figure~\ref{fig1} shows the beam energy and $p_{\mathrm T}$
dependence of $R_{\mathrm {CP}}$ for $\pi^{+}$+$\pi^{-}$ and
$p$+$\bar{p}$ in Au+Au collisions.
 There is a distinct difference in the $p_{\mathrm T}$ dependence
 of $R_{\mathrm {CP}}$ for charged pions and protons+anti-protons 
 observed at both energies. 
 The $R_{\mathrm {CP}}$ values are higher at $\sqrt{s_{\mathrm {NN}}}$ =~62.4~GeV
 than for $\sqrt{s_{\mathrm {NN}}}$ = 200 GeV
 for  $p_{\mathrm T}$ $<$ 7 GeV/c; beyond this $p_{\mathrm T}$
 they seem to approach each other. 
 Considering a lower parton energy loss
 due to a smaller initial gluon density at the lower energy, one may
 expect the $R_{\mathrm {CP}}$ to approach each other at high
 $p_{\mathrm T}$ due to a steeper initial jet spectra at 62.4
GeV. At high $p_{\mathrm T}$ a similar value of 
 $R_{\mathrm {CP}}$ for $p$+$\bar{p}$ and $\pi^{+}+\pi^{-}$ 
is observed at both energies.
This may indicate a similar energy loss
of quarks and gluons in the medium formed in high energy
heavy ion collisions.

\subsection{Baryon to Meson Ratio}
\begin{figure}
\begin{center}
\includegraphics[scale=0.28]{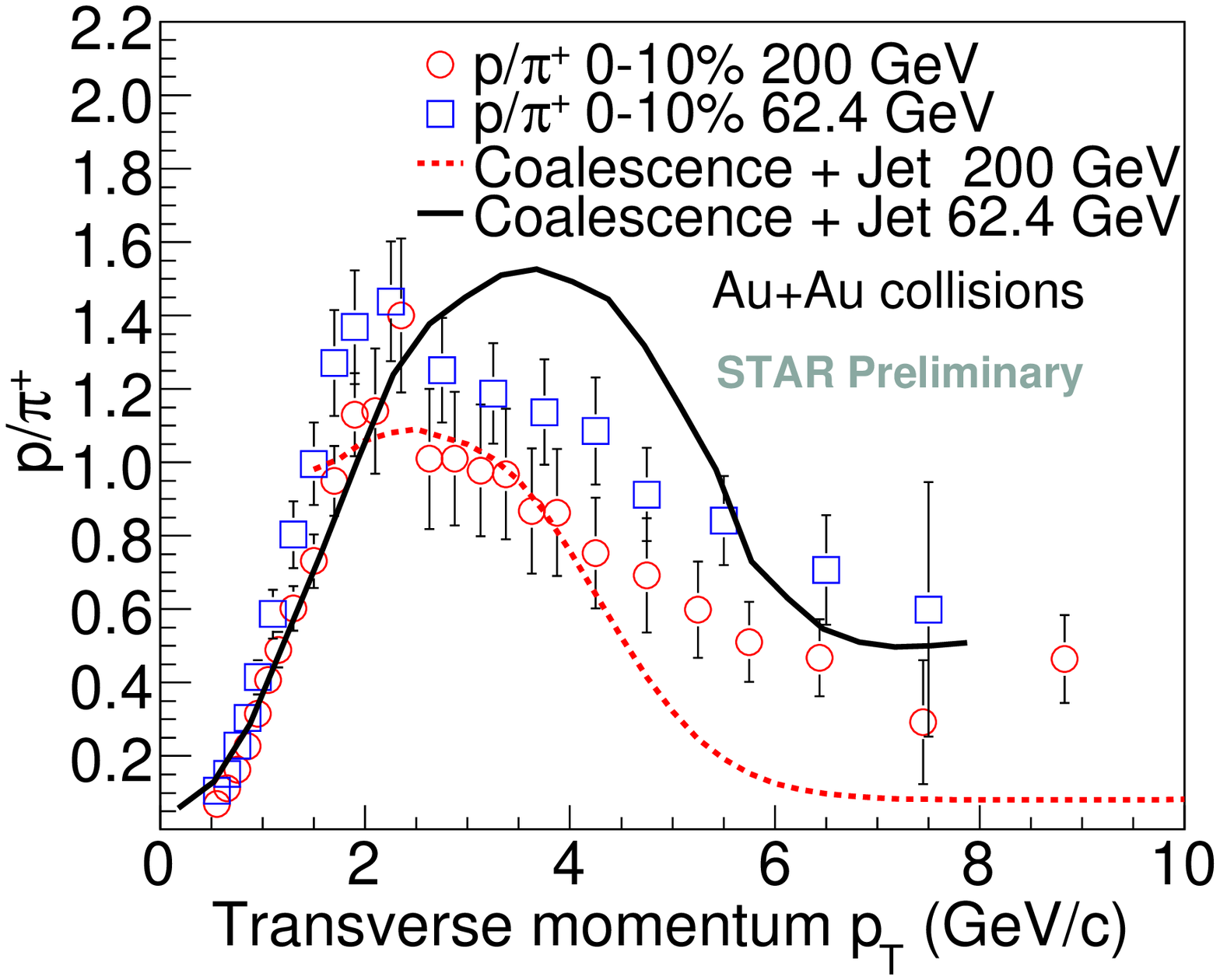}
\includegraphics[scale=0.28]{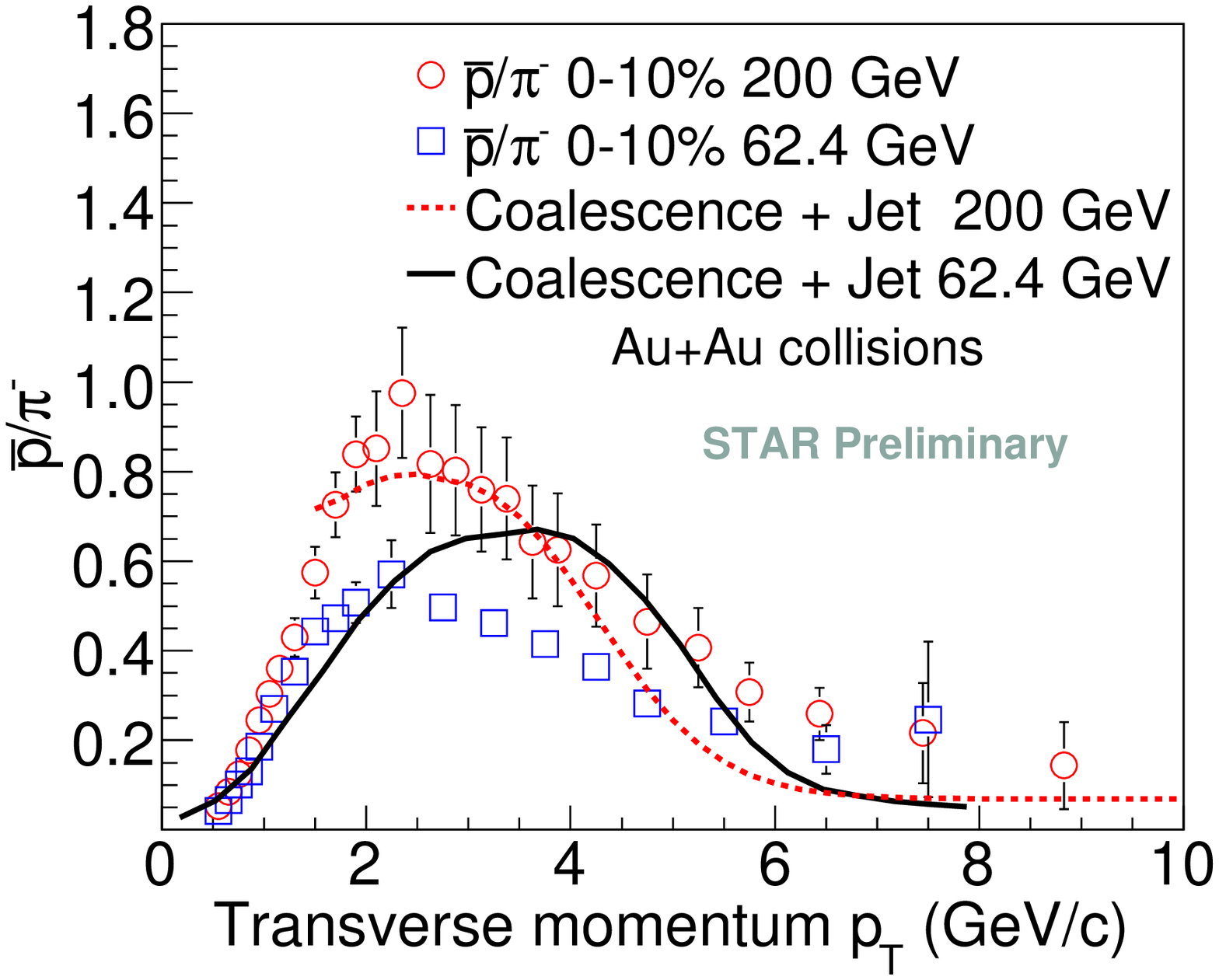}
\caption{$p/\pi^{+}$ (left panel) and $\bar{p}/\pi^{-}$ (right panel) 
for 0-10\% Au+Au collisions at 
$\sqrt{s_{\mathrm {NN}}}$ = 200 and 62.4 GeV. The error bars are
systematic and statistical errors added in quadrature. 
Data are compared to model calculations based on coalescence and
jet fragmentation in Refs.~\cite{vitev_62,fries}.}
\label{fig2}
\end{center}
\end{figure}

Figure~\ref{fig2} shows the $p/\pi^{+}$ and $\bar{p}/\pi^{-}$ ratios
for Au+Au collisions at $\sqrt{s_{\mathrm {NN}}}$~=~200 and 62.4~GeV.
The experimental results are compared to model calculations based
on coalescence and jet fragmentation~\cite{vitev_62,fries}.
The $p/\pi^{+}$ ratio for Au+Au collisions at 
$\sqrt{s_{\mathrm {NN}}}$~=~62.4~GeV is higher than the
corresponding values at $\sqrt{s_{\mathrm {NN}}}$ = 200 GeV
in the intermediate $p_{\mathrm T}$ range.
The case for $\bar{p}$/$\pi^{-}$ ratio is reversed. The 
higher value of $p/\pi^{+}$ at intermediate $p_{\mathrm T}$ in 62.4 GeV 
compared to 200 GeV is consistent with quark coalescence models~\cite{vitev_62}.
However a detailed comparison of the $p/\pi^{+}$ and $\bar{p}/\pi^{-}$ ratios
to predictions from model shows there is a lack of quantitative agreement.
A shift in the peak position for the ratios at 62.4 GeV predicted by the 
models are not observed in the data.

\subsection{System size dependence}

\begin{figure}
\begin{center}
\includegraphics[scale=0.38]{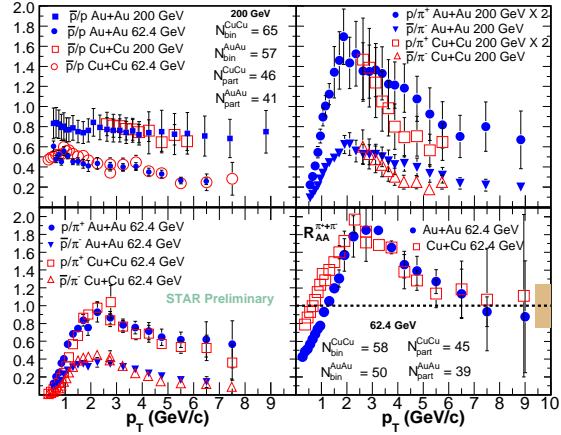}
\caption{
Comparison of $\bar{p}/p$ (top left panel) , $p/\pi^{+}$, $\bar{p}/\pi^{-}$ (top right
and bottom left panels) and 
$R_{AA}^{\pi^{+}+\pi^{-}}$ (bottom right panel)  
for Au+Au and Cu+Cu collisions having similar number of
participating nucleons ($N_{\mathrm {part}}$) and number of binary
collisions  ($N_{\mathrm {bin}}$) at 
$\sqrt{s_{\mathrm {NN}}}$ = 200 and 62.4 GeV. The error bars are systematic 
and statistical errors added in quadrature. The shaded band in the
panel showing $R_{AA}$ is the uncertainty due to $N_{\mathrm {bin}}$
calculations. $R_{AA}$ has an additional 25\% uncertainty due to the parametrization
used for p+p data at $\sqrt{s_{\mathrm {NN}}}$ = 62.4 GeV~\cite{david}.}
\label{fig3}
\end{center}
\end{figure}

Figure~\ref{fig3} shows the comparison of $\bar{p}/p$,
$p/\pi^{+}$, $\bar{p}/\pi^{-}$ and 
$R_{AA}^{\pi^{+}+\pi^{-}}$ ($=\,\frac{d^2N_{\rm{AuAu}}/dy dp_{\rm T}}{(\langle N_{\rm
{bin}}\rangle /\sigma_{\rm{pp}}^{\rm
{inel}})\,d^2\sigma_{\rm{pp}}/dy dp_{\rm T}}$) for Au+Au and Cu+Cu
collisions having similar $N_{\mathrm {part}}$ and $N_{\mathrm {bin}}$
at $\sqrt{s_{\mathrm {NN}}}$ = 200 and 62.4 GeV. For the $R_{AA}^{\pi^{+}+\pi^{-}}$ 
at $\sqrt{s_{\mathrm {NN}}}$ = 62.4 GeV, the 
$\sigma_{\rm{pp}}^{\rm
{inel}}$ is taken to be 36 mb and the $p+p$ data are from the 
parametrization available from ISR data in the Ref.\cite{david}. 
We observe, for similar $N_{\mathrm {part}}$ and/or $N_{\mathrm {bin}}$
at a given beam energy, the above observables at high $p_{\mathrm T}$  have
similar values in Au+Au and Cu+Cu collisions.

\section{Summary}

We have presented a study of the energy dependence of the 
$\pi^{\pm}$, $p$ and $\bar{p}$ production at high $p_{\mathrm T}$ 
from Au+Au and Cu+Cu collisions at 
$\sqrt{s_{\mathrm {NN}}}$ = 62.4 and 200 GeV.  
There is a distinct difference in the $p_{\mathrm T}$ dependence 
of $R_{\mathrm CP}$ for charged pions and protons+anti-protons 
observed at both energies. However at higher $p_{\mathrm T}$ 
the values of $R_{\mathrm {CP}}$ for baryons and 
mesons at both energies are similar. This together  
with the recent observation
of a comparable $\bar{p}/\pi^{-}$ ratio in 200 GeV Au+Au collisions
and $d$+Au collisions~\cite{star_pid200,ppdau} at high $p_{\mathrm T}$ 
provides data necessary to understand the differences, if any, in the interaction of quarks and
 gluons and the medium, and through this the energy loss mechanism. The $p$/$\pi^{+}$ ratio 
for Au+Au collisions at $\sqrt{s_{\mathrm {NN}}}$ = 62.4 GeV 
is higher than corresponding values at $\sqrt{s_{\mathrm {NN}}}$ = 200 GeV 
in the intermediate  $p_{\mathrm T}$ range. 
The case for $\bar{p}$/$\pi^{-}$ ratio is reversed. The higher 
value of $p$/$\pi^{+}$ ratio at intermediate $p_{\mathrm T}$ 
for 62.4 GeV compared to 200 GeV is 
qualitatively consistent with quark coalescence models.
 However there are quantitative disagreements 
of data with such models. We also observe similar values
of high $p_{\mathrm T}$ particle ratios and $R_{AA}$ in Au+Au and Cu+Cu collisions 
for collisions with similar $N_{\mathrm {part}}$ and/or $N_{\mathrm {bin}}$.

\end{document}